\documentclass[twocolumn,showpacs,preprintnumbers,amsmath,amssymb]{revtex4}
\usepackage{graphicx}
\usepackage{dcolumn}
\usepackage{bm}
\usepackage[latin1]{inputenc}

\usepackage{color}

\begin{document}

\title{Cavity-controlled collective scattering at the recoil limit}

\author{Simone Bux$^a$}
\author{Christine Gnahm$^a$}
\author{Reinhardt A.W. Maier$^a$}
\author{Claus Zimmermann$^a$}
\author{Philippe W. Courteille$^{ab}$}
\affiliation{$^a$Physikalisches Institut, Eberhard-Karls-Universit\"at T\"ubingen, Auf der Morgenstelle 14, D-72076 T\"ubingen, Germany.\\
	$^b$Instituto de F\'isica de S\~ao Carlos, Universidade de S\~ao Paulo, 13560-970 S\~ao Carlos, SP, Brazil.}

\date{\today}

\begin{abstract}
We study collective scattering with Bose-Einstein condensates interacting with a high-finesse ring cavity. The condensate scatters the light of a transverse pump beam superradiantly into modes which, in contrast to previous experiments, are not determined by the geometrical shape of the condensate, but specified by a resonant cavity mode. Moreover, since the recoil-shifted frequency of the scattered light depends on the initial momentum of the scattered fraction of the condensate, we show that it is possible to employ the good resolution of the cavity as a filter selecting particular quantized momentum states.
\end{abstract}


\maketitle

Under certain circumstances optical and matter wave modes can interact on equal footings in a four-wave mixing process \cite{Law98}. Recent examples for this are the observations of light-induced collective instabilities in cold atomic clouds \cite{Inouye99,Kozuma99}. The instabilities are induced by mutual Bragg scattering of light at a matter wave grating and atoms at an optical standing wave. The scattering takes place as a self-amplified process called matter wave superradiance (MWSR) or collective atomic recoil lasing (CARL) depending on how subsequent scattering events are correlated. In the case of MWSR, the correlations are stored in long-lived matter wave interferences developing in an ultracold cloud \cite{Inouye99}. In general, the scattered photons rapidly leave the interaction volume, thus limiting the coherence time of the optical mode. In the case of CARL, the decay is controlled by recycling the scattered photons in a high-finesse ring cavity. As a consequence, the correlations between scattering events can also be stored in long-lived optical modes of the cavity \cite{Slama07,Slama07b}.


The interaction of ultracold atoms with optical cavities has been studied in several experiments \cite{Colombe07,Brennecke07,Baumann10} aiming at reaching the strong coupling limit, where cavity quantum electrodynamics (CQED) can be studied with Bose-Einstein condensed (BEC) atomic clouds. Coupling strengths exceeding not only the cavity decay rate, but also the natural decay rate of the excited atomic state are achieved with microcavities. The mode volumes of these cavities are small enough for a single photon to produce a field strength saturating the atomic transition. However, a small mode volume necessarily implies a poor spectral resolution of the cavity.

In this Letter, using a large ring cavity (round trip length $L=87\,$mm) with a very high finesse of $F=135000$, we address the opposite regime characterized by an extremely high resolution on the order of the recoil frequency. At the same time, we maintain the \emph{collective} coupling strong. A BEC located inside the cavity is illuminated from the side  with a pump laser pulse. Using collective scattering in a combination of MWSR and CARL as a probe, we demonstrate the cavity's dramatic impact on the light scattering in two ways. First, we show that the cavity is able to lift the superradiant gain above threshold provided it is resonant with the scattered light. The cavity frames the direction of the superradiant modes, although the symmetry axis of the elongated BEC considerably deviates from the cavity's optical axis. Second, the atomic momentum distribution shows a very sensitive dependence on the pump laser frequency. A modification of only $20\,$kHz completely alters the momentum distribution, such that, in principle, it is possible to address and populate selected quantized momentum states. The control over the atomic motion in the scattering process represents a further step towards the experimental implementation of a light-matter interaction, where \emph{all} involved degrees of freedom are free from dissipation and whose excitations are frozen out or quantized.

Our experiment is very similar to the original matter wave superradiance experiment \cite{Inouye99}, where a short pump laser pulse (intensity $I$, detuning $\Delta_a$) irradiated into a BEC gave rise to a pattern of recoiling atoms coupled out of the condensate, while at the same time a superradiant burst of light was emitted into the long axis of the condensate. The rate $R_{SR}$ at which an ensemble of $N$ atoms cooperatively scatters light into a solid angle $\Omega_{sol}$ was found to be
\begin{equation}\label{Eq_SRrate}
	R_{SR}=R_1~N~\tfrac{N_r+1}{2}~,
\end{equation}
where $R_1=\sin^2\theta\cdot\sigma(\Delta_a)\cdot(I/\hbar\omega)\cdot(3\Omega_{sol}/8\pi)$ is the single-atom scattering rate. $\sigma(\Delta_a)$ is the off-resonant optical cross section and $\theta$ is the angle between the pump light polarization and the direction into which the light is scattered. For elongated BECs the solid angle may be approximated by $\Omega_{sol}\simeq(2\eta)^{-2}$, where $\eta$ is the aspect ratio. $N_r$ is the number of atoms already populating the recoil mode, into which the atoms $N$ are pumped when scattering photons into the solid angle.

In our experiment, the presence of a cavity breaks the isotropy of the density of optical modes capable of receiving scattered photons. The solid angle covered by the cavity, $\Omega_{sol}=8\pi/(kw_0)^2$, where $k$ is the wavenumber and $w_0\simeq100\,\mu$m the waist of the Gaussian mode, is very small. Hence, despite the high finesse, the cavity-to-free space scattering ratio (Purcell factor) of our cavity is small, $2\Omega_{sol}F/\pi=0.4$, meaning that the natural decay rate of the atom is not considerably altered. On the other hand, the cavity profoundly alters the scattering rate $R_{cv}$ into the cavity mode \cite{Heinzen87}, $R_{cv}=\mathcal{L}(\Delta_c)R_{SR}$, where $\mathcal{L}(\Delta_c)\equiv\frac{\sqrt{1+(2F/\pi)^2}}{1+(2F/\pi)^2\sin^2(\Delta_c/\delta_{fsr})}~$, with the detuning from the cavity mode $\Delta_c$, the free spectral range $\delta_{fsr}=\kappa F/\pi$ and the cavity decay width $\kappa=(2\pi)\cdot12.7\,$kHz. If the cavity is non-resonant, $\mathcal{L}(\omega)\simeq0$, no photons are scattered into the mode volume. In the resonant case, the enhancement factor is $2F/\pi$. In other words, we expect that the cavity influences the scattering process not via a superradiant modification of the rate at which photons are scattered, but via a reorientation of the scattering direction. An important point is, that $\kappa$ is smaller than the maximum recoil shift $\omega_{rec}\equiv2\hbar k^2/m_{Rb}=(2\pi)\cdot14.5\,$kHz, so that we expect sub-recoil sensitivity of the scattering dynamics on the pump frequency.

\begin{figure}[b!]
		\centerline{\scalebox{.43}{\includegraphics{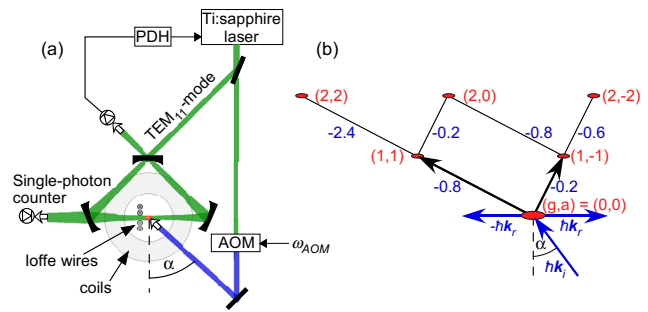}}}
		\caption{(color online) (a) Geometry of the experiment: A BEC is created in a Ioffe-Pritchard trap.
			The BEC is placed in the waist of a TEM$_{11}$-mode of an optical ring cavity (green).
			A pump beam is irradiated from the side under the angle $\alpha=37^\circ$. A single-photon counter records the photons transmitted through one of the cavity mirrors.
			(b) Momentum distribution into which the BEC evolves after interaction with the pump beam. The momentum states are labeled with red indices $(g,a)$, 
			the explanation is given in the text. The blue numbers denote the frequency shift of the scattered photons $\Delta\omega_{g,a,\pm}$ 
			(in units of $\omega_{rec}=(2\pi) \cdot 14.5\,$kHz) [see Eq.~(\ref{Eq_scatteredfrequency})].}
		\label{Fig1_Scheme}
	\end{figure}

Our experimental setup for preparing a $^{87}$Rb Bose-Einstein condensate in a Ioffe-Pritchard (IP) type magnetic trap in a high-finesse ring cavity has been detailed in Refs.~\cite{Slama07,Slama07b}. The minimum of the IP trapping potential is located slightly outside the mode volume of the cavity. After quantum degeneracy has been reached, the BEC is transported to the cavity mode by displacing the center of the IP trap. The BEC is now illuminated by $s$-polarized pump light incident under an angle of $\alpha=37^\circ$ with respect to a normal on the cavity's optical axis and within the cavity's plane [see Fig.~\ref{Fig1_Scheme}(a)]. A Ti:sapphire laser providing the pump beam of frequency $\omega_{pp}$ is detuned by an amount $\Delta_a=\omega_{pp}-\omega_{D1}$ relative to the rubidium $D_1$ resonance line ($F=2 \rightarrow F'=2$). In order to precisely control the detuning $\Delta_c$ of the pump with respect to a TEM$_{00}$ cavity resonance, $\Delta_c=\omega_{pp}-\omega_{TEM00}$, resonant light is additionally injected collinearly into the cavity. We minimize the impact of this reference light on the atomic cloud by two measures. Firstly, the reference light is injected with low power on a TEM$_{11}$-mode of the cavity exhibiting an intensity minimum near the optical axis, where the atomic cloud is located \cite{Bux07}. The pump beam frequency is shifted from the reference light by means of an acousto-optic modulator (AOM) in order to lie close to a TEM$_{00}$-resonance. Without atoms in the mode volume, the frequency separation of the cavity modes is $\omega_{TEM00}-\omega_{TEM11}=-163.37\,$MHz.
The second measure consists in providing the laser beams only at the last moment, shortly before they are needed in the pulse sequence. Directly after the atoms have reached their final position inside the cavity, a sample-and-hold circuitry activates a Pound-Drever-Hall servo, which locks the reference laser to the TEM$_{11}$-mode. A few microseconds later, a pulse is delivered to the AOM controlling the pump beam. The duration $\tau$ of the pulse, the pump beam frequency $\omega_{pp}$ and the intensity $I$ can be tuned. At the end of the pulse, the laser and the IP trap are switched off instantly. The atoms fall freely during $15\,$ms before their momentum distribution is mapped via absorption imaging. Additionally, we record the photons scattered into the cavity and transmitted through one of the cavity mirrors [see Fig.~\ref{Fig1_Scheme}(a)].

\begin{figure}[b!]
		\centerline{\scalebox{.195}{\includegraphics{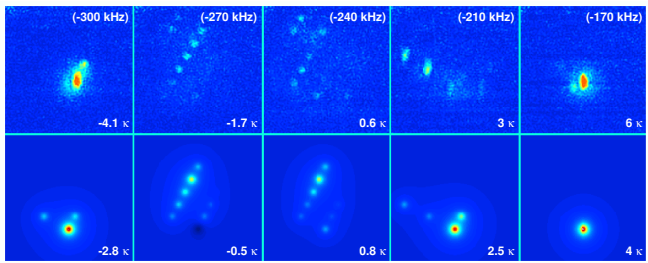}}}
		\caption{(color online) Momentum distributions of $N=80000$ atoms observed with different pump laser detunings $\Delta_c$ (in brackets) after a $\tau=200\,\mu$s 
			pump laser pulse.
			The detunings relative to the cavity resonance shifted by the atoms in units of $\kappa$ are evaluated from Lorentzian fits to the scattered photon number
			(explanation in the text). The upper row corresponds to $\Delta_a=-(2\pi)\cdot4.7\,$GHz and $I=50\,\text{mW/cm}^2$, the lower row shows simulated momentum
			distributions based on the model Eq.~(\ref{Eq_rate}). The simulations assume the same experimental parameters, except for a $5$ times higher intensity.}
		\label{Fig2_SpectrumTOF}
	\end{figure}

Within a narrow range of detuning $\Delta_c$ between the pump frequency and the cavity resonance, and only for $s$-polarized pump light, we observe superradiant scattering. The signatures are a characteristic fan of atomic momentum states, as shown in Fig.~\ref{Fig2_SpectrumTOF}, and counts of photons leaking through the cavity mirrors. The shape of the momentum distribution and the number of scattered photons critically depend on $\Delta_c$. If we vary the pump laser frequency by only a few $\kappa$, we observe a different momentum distribution. The reason for this is that the scattered light  
experiences frequency-shifts depending on the velocity of the scattering momentum state and on the direction into which the light is scattered (left- or right-going cavity mode). The degeneracy of the scattering direction is broken by the angle of incidence, which is nonzero. The extreme resolving power of the cavity segregates between the scattered frequency components, only amplifying the resonant ones. Consequently, only those atomic states are scattering, which have the right velocity component along the direction of the total photonic recoil.

The scattering of a photon with frequency $\omega_i$ from the pump mode $\mathbf{k}_i$ into the mode $\mathbf{k}_r$ by an atom initially moving with momentum $\mathbf{p}_i$ produces an atom in the momentum state $\mathbf{p}_r$. Note that a right-going photon produces a left-going atom and vice versa due to momentum conservation. The \emph{initial} momentum of all atoms is $\mathbf{p}_i=0$. However, atoms may scatter several photons, i.e.~the \emph{instantaneous} $\mathbf{p}_i$ depend on how many photons have been scattered previously. Atoms having scattered several photons give rise to further generations of momentum states that are labeled with the number $g$ of photons scattered. The fact that photons can only be scattered into the two cavity modes greatly simplifies the situation, since momentum states having scattered the same ratio of photons into the left- and into the right-going cavity mode are practically degenerate. It is thus sufficient to introduce as a second label the difference $a$ in number of left- and right-going photons and obtain the following momenta for the atomic momentum states,
\begin{equation}\label{Eq_momenta}
	\mathbf{p}_r \equiv \mathbf{p}_{g,a}=g\hbar\mathbf{k}_i-a\hbar\mathbf{k}_r~,
\end{equation}
for all $a=-g,-g+2,..,g$. Furthermore, we label the photons scattered into the left and right-going cavity modes by ($-$) and ($+$), respectively. Eliminating $\mathbf{p}_r$ from the conservation equations for momentum and energy and using $\hbar^2k_i^2/2m\approx\hbar^2k_r^2/2m$, we obtain the spectrum of the light scattered into the cavity as a function of the atomic momentum before the scattering process,
\begin{equation}
	\Delta\omega_{g,a,\pm}=\tfrac{\omega_{rec}}{2}\left(-1\pm\tfrac{\mathbf{k}_r\mathbf{k}_i}{k^2}\right)+\tfrac{\mathbf{p}_{g-1,a}}{m}(\pm\mathbf{k}_r-\mathbf{k}_i)~.
\end{equation}
The first term describes photonic recoil, the second accounts for the Doppler shift. Substituting Eq.~(\ref{Eq_momenta}), we obtain
\begin{eqnarray}\label{Eq_scatteredfrequency}
	\Delta\omega_{g,a,\pm}=\tfrac{\omega_{rec}}{2}(g\pm a)(-1\pm\sin\alpha)~.
\end{eqnarray}
A scheme of the momentum distribution and the calculated frequencies $\Delta\omega_{g,a,\pm}$ is plotted in Fig.~\ref{Fig1_Scheme}(b). One can see, that scattering of atoms to the right leads to a smaller frequency shift than scattering to the left.

Obviously, the frequencies of the scattered photons not only depend on the direction into which they are scattered, but also at which momentum mode $(g,a)$ they are scattered. The Doppler shift thus introduces a whole spectrum of frequencies. Only the resonant ones, $|\Delta\omega_{g,a,\pm}|<\kappa$, are supported and eventually amplified by the cavity. Scattering from higher momentum states giving rise to large Doppler shifts is disrupted.

Since each momentum state can scatter to the left or to the right, two scattering processes may contribute to every momentum state. This may be expressed by the following rate equation for the atom numbers $N_{g,a}$ in the states $(g,a)$, keeping in mind that $N_{g,a}=0$ for $g<|a|$ or $g<0$,
\begin{align}\label{Eq_rate}
    \dot N_{g,a} & = \sum_\pm\left(R_{g-1,a\pm1,\mp}~N_{g-1,a\pm1}\tfrac{N_{g,a}+1}{2}\right.\\
    	& \left. -R_{g,a,\pm}~N_{g,a}\tfrac{N_{g+1,a\pm1}+1}{2}\right)~,\nonumber
\end{align}
with $R_{g,a,\pm}=\mathcal{L}(\Delta\omega_{g,a,\pm})R_1$ and the single-atom Rayleigh scattering rate $R_1$ from Eq.~(\ref{Eq_SRrate}).

The atomic momentum states generated by the scattering of light into the cavity give rise to a matter wave grating oriented such as to encourage subsequent photons to be scattered into the cavity as well. This dynamics is described by our simple rate equation model (\ref{Eq_rate}). However, at the same time, the photons scattered into the ring cavity are recycled into the interaction zone. There they form together with the incident pump light an optical lattice further amplifying the scattering dynamics. This CARL type amplification is not contained in our simple model. Nevertheless, Eq. (6) is able to simulate momentum distributions which are qualitatively similar to the observed ones, as one can see in the lower row of Fig.~\ref{Fig2_SpectrumTOF}. We performed complementary measurements of the temporal evolution of the populations $N_{g,a}$ showing that superradiant enhancement of the gain indeed plays a role, but it is not possible at the present stage to quantify the role of CARL. A full quantum treatment of our system will be the topic of future work.
	\begin{figure}[t!]
		\centerline{\scalebox{.43}{\includegraphics{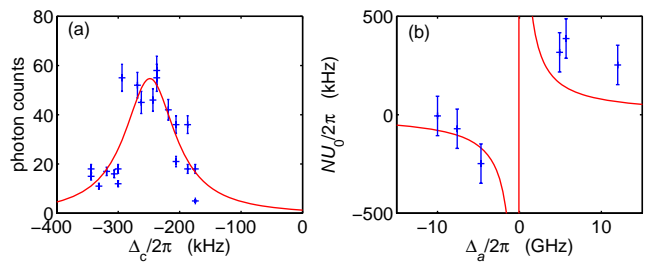}}}
		\caption{(color online) (a) Dependence of the scattered photon number recorded within $\tau=200\,\mu$s on $\Delta_c$. The parameters are the same as in
			Fig.~\ref{Fig2_SpectrumTOF}. The errorbars account for a $10\%$ estimated uncertainty in the number of counted photons. The solid line shows a Lorentzian fit.
			(b) Dependence of cavity mode shift $U_N$ on $\Delta_a$. The solid line shows a calculation of $U_N$. Note that the trapped hyperfine ground state connects via
			$\pi$-light only to the same hyperfine state in the $^2P_{1/2}$ multiplet. The blue crosses denote the detuning of maximum scattering, $\Delta_{c,m}$.
			The errorbars account for a $10\%$ estimated error.} 
		\label{Fig3_CavityDispersion}
	\end{figure}
	
\begin{figure}[b!]
		\centerline{\scalebox{.44}{\includegraphics{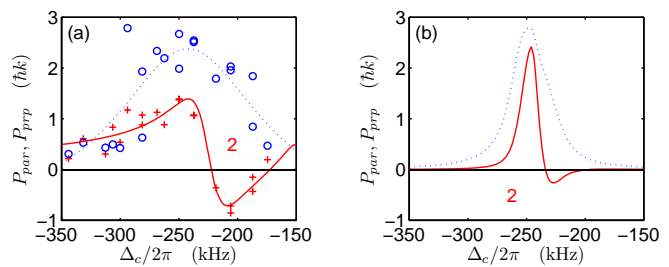}}}
		\caption{(color online) Perpendicular (red crosses/line) and longitudinal (blue circles/dotted line) projection of the mean momentum on the pump direction.
			(a) Measurement of the dependence on the AOM-frequency. The parameters are the same as in the upper row of Fig.~\ref{Fig2_SpectrumTOF}.
			The lines are to guide the eye. (b) Simulation based on the rate equation (\ref{Eq_rate}). The resonance shift $U_N$ was calculated with $N=80000$.}
		\label{Fig4_DispersiveCurve}
	\end{figure}

The photons transmitted through one of the cavity mirrors (transmission $T=1.5$ppm) are detected with a single-photon counter (quantum efficiency $\eta=30\%$) [see Fig.~\ref{Fig1_Scheme}(a)]. Under the experimental conditions, we expect to count roughly $\eta TR_{cv}F/\pi\simeq20$ photons scattered into a resonant cavity mode per $\tau=200\,\mu$s. This agrees with the photon number recorded as a function of $\Delta_c$ as shown in Fig.~\ref{Fig3_CavityDispersion}(a). The exhibited resonance curve is shifted by a value depending on the detuning $\Delta_a$. The reason for this is that the refraction index of the atomic cloud occupying the mode volume shifts the resonance frequency of the TEM$_{00}$-mode with respect to the TEM$_{11}$-resonance by an amount corresponding to the collective single-photon light shift $U_N\equiv N\Omega_1^2\Delta_a/(\Delta_a^2+\kappa^2)$, where $\Omega_1=(2\pi)\cdot120\,$kHz is the single-photon Rabi frequency of our cavity. From Lorentzian fits to the resonance profile we determine the detuning $\Delta_{c,m}$ of maximum scattering for different $\Delta_a$. Fig.~\ref{Fig3_CavityDispersion}(b) shows, that the $\Delta_a$-dependence of $\Delta_{c,m}$ is dispersive and agrees reasonably with the one of $U_N$. Moreover, the resonance curve Fig.~\ref{Fig3_CavityDispersion}(a) is clearly broadened with respect to the empty cavity transmission spectrum. This is due, on one hand, to shot-to-shot fluctuations in the atom number (about $10\,\%$) leading to a fluctuating resonance shift $U_N$ \cite{NoteBadSumming}. On the other hand, photons in the cavity mode may be scattered out of the cavity at the atomic cloud, which represents a loss channel reducing the finesse. 
Via the pump detuning $\Delta_c$, we can control the main scattering direction. Fig.~\ref{Fig4_DispersiveCurve}(a) shows the longitudinal and perpendicular projection of the measured momentum distribution onto the pump beam direction. The perpendicular projection exhibits a dispersive dependence on detuning. Interestingly, the simulation of the rate equations shows the same tendency [Fig.~\ref{Fig4_DispersiveCurve}(b)]. The measured curve is broader for the same reasons which broaden the resonance in Fig.~\ref{Fig3_CavityDispersion}(a). The dispersive behaviour can be understood directly from Fig.~\ref{Fig2_SpectrumTOF}, as for negative detuning predominantly momentum states on the left of the pump direction are occupied, whereas for positive detuning those on the right are populated. Therefore the orthogonal projection goes from positive to negative as the detuning is increased.

In conclusion, we demonstrated that an optical cavity can be used as a macroscopic handle for the least quantized, but still relevant degree of freedom in the light-matter interaction, i.e.~the atomic center-of-mass motion. The mere fact that the resonant cavity sucks off the scattered photons leads to a modification of the force accelerating the atoms. Due to its very high resolution the cavity is capable of filtering quantized momentum states, simply because they Doppler-shift the scattered light by different amounts and the cavity selectively amplifies the resonant momentum transitions.

The method of using the cavity as a filter may allow for trapping excited atomic velocity states, which is interesting for resolved-sideband cavity cooling \cite{Elsaesser03}. Even a controlled generation of superpositions of momentum states may be possible using time-dependent pulse sequences, multi-component or phase-modulated pump beams \cite{Cola09}.

This work has been supported by the Deutsche Forschungsgemeinschaft (DFG) and by the Carl-Zeiss-Stiftung.


\begin{thebibliography}{10}
\bibitem{NoteBadSumming}Clearly, the method of summing over the contributions of all frequencies, $\sum_{g,a}\mathcal{L}(\Delta\omega_{g,a,\pm})$, to the number of photons scattered into the cavity is an approximation. Furthermore, the development of a matter wave lattice can modify the frequency shift $U_N$ in time.
\end{thebibliography}
\end{document}